\documentclass[cits]{PoS}\usepackage{amsmath,cite,color,bm,booktabs,dcolumn}

\newcolumntype{d}[1]{D{.}{.}{#1}}
\title{Isospin Breaking in Heavy-Meson Decay Constants}
\ShortTitle{Isospin Breaking in Heavy-Meson Decay Constants}
\author{Wolfgang Lucha\\Institute for High Energy Physics,
Austrian Academy of Sciences, Nikolsdorfergasse 18, A-1050 Vienna,
Austria\\E-mail: \email{Wolfgang.Lucha@oeaw.ac.at}}
\author{\speaker{Dmitri Melikhov}\\Institute for High Energy Physics,
Austrian Academy of Sciences, Nikolsdorfergasse 18, A-1050 Vienna,
Austria, and\\ D.~V.~Skobeltsyn Institute of Nuclear Physics,
M.~V.~Lomonosov Moscow State University, 119991, Moscow, Russia,
and\\ Faculty of Physics, University of Vienna, Boltzmanngasse 5,
A-1090 Vienna, Austria\\E-mail: \email{dmitri\_melikhov@gmx.de}}
\author{Silvano Simula\\INFN, Sezione di Roma Tre, Via della Vasca
Navale 84, I-00146 Roma, Italy\\E-mail:
\email{simula@roma3.infn.it}} 

\abstract{Evaluation of Borelized QCD sum rules in the so-called
local-duality limit of infinitely large Borel mass parameter
provides an alternate route for extraction of the dependence of
the decay constants of heavy--light mesons on the mass $m_q$ of
the involved light quark $q$: For appropriate choices of the
two-point correlation functions of currents interpolating the
hadrons under study, the local-duality limit forces all
nonperturbative contributions parametrized by vacuum condensates
to such kind of correlator to vanish. As a consequence, the sought
$m_q$ dependence of the heavy--light meson decay constants proves
to be controlled \emph{primarily\/} by the correlator
contributions from perturbative~QCD. Our knowledge of the analytic
behaviour of the latter as functions of $m_q$ enables us to derive
the $m_q$ dependence of the decay constants of both pseudoscalar
and vector heavy--light mesons,~for which we estimate strong
isospin breaking to be of the order of $1\;\mbox{MeV}$ for both
charm and beauty~sectors.}

\FullConference{XVII International Conference on Hadron
Spectroscopy and Structure --- Hadron2017\\25--29 September,
2017\\University of Salamanca, Salamanca, Spain}\begin{document}

\section{Local-Duality Limit of QCD Sum Rules for Heavy--Light
Meson Decay Constants}QCD sum rules \cite{QSR} constitute a
nonperturbative approach to bound states of quarks and gluons, the
degrees of freedom of quantum chromodynamics (QCD), the quantum
field theory governing all strong interactions. This kind of
relations may be constructed by evaluating, at the level of QCD
and at the hadronic level, correlation functions of operators
appropriately defined in terms of quarks~and gluons but
interpolating the hadron of interest. Application of Borel
transformations from momenta to new variables, the Borel
parameters $\tau,$ suppresses (unwanted) hadronic-continuum
contributions. For heavy--light mesons $H_q$ of mass $M_{H_q}$
composed of a heavy quark $Q=c,b$ of mass $m_Q$ and a light quark
$q=u,d,s$ of mass $m_q$, the Borel QCD sum rules for their decay
constants $f_{H_q}$ generically~read
$$f_{H_q}^2\left(M_{H_q}^2\right)^N\exp(-M_{H_q}^2\,\tau)=
\hspace{-3.224ex}\int\limits_{(m_Q+m_q)^2}^{s^{(N)}_{\rm
eff}(\tau,m_Q,m_q,\alpha_{\rm s})}\hspace{-3.224ex}{\rm
d}s\exp(-s\,\tau)\,s^N\,\rho(s,m_Q,m_q,\alpha_{\rm
s})+\Pi^{(N)}(\tau,m_Q,m_q,\alpha_{\rm s},\langle\bar
qq\rangle,\dots)\ ,$$with a positive integer exponent
$N=0,1,\dots$ fixed by the detailed formulation of the QCD sum
rule. The spectral densities $\rho(s,m_Q,m_q,\alpha_{\rm s})$ can
be found in form of expansions in the strong~coupling $\alpha_{\rm
s}$. The $\tau$-dependent effective threshold $s^{(N)}_{\rm
eff}(\tau,m_Q,m_q,\alpha_{\rm s})$ forms the lower boundary of
that region of $s$ (extending to infinity) over which, by the
postulate of quark--hadron duality, mutual cancellations of the
contributions of perturbative QCD and of hadronic excitations and
continuum should take place. Basically, nonperturbative effects
manifest in QCD sum rules in two places: as vacuum condensates in
power corrections $\Pi^{(N)}(\tau,m_Q,m_q,\alpha_{\rm
s},\langle\bar qq\rangle,\dots)$, power series in $\tau$, and in
$s^{(N)}_{\rm eff}$. Depending on that (in fact, chosen) number
$N$, their relative fractions in power corrections and effective
threshold~vary.

The conventional procedure of deriving, from such QCD sum rule,
the sought relation between the hadron characteristics of interest
and the fundamental parameters of QCD starts by identifying a
suitable interval of (thus inevitably almost everywhere nonzero)
values of $\tau$, defined such that, at the hadron side, the
ground-state contribution is reasonably large and, at the QCD
side, nonperturbative corrections stay sufficiently small.
Equipped with the increase of the accuracy \cite{LMSAU} of the
predictions gained by taking seriously the $\tau$ dependence
\cite{LMSET} of the effective threshold, and determining the
latter by minimizing the discrepancy between theoretical hadron
masses and their true values known from experiment, we managed to
extract precise decay-constant predictions \cite{LMSDC,LMSR} from
the required set of QCD quantities, such as quark masses, strong
coupling, spectral densities and vacuum condensates.

The --- compared to typical hadron masses tiny --- difference
$(m_d-m_u)(2\;\mbox{GeV})\approx2.5\;\mbox{MeV}$ \cite{PDG} of the
down-quark mass $m_d$ and the up-quark mass $m_u$ generates
strong-isospin breakdown reflected by the difference
$f_{H_d}-f_{H_u}$ of the decay constants of heavy--light mesons
$H_d$ and $H_u$ involving, apart from a given heavy quark, a light
$d$ and $u$ quark, respectively. Our QCD sum-rule version relying
on $\tau$-dependent thresholds proves to be a tool so sharp that
we can reliably treat such phenomena \cite{LMScf,LMSIB}.

\begin{figure}[h]\begin{center}\begin{tabular}{ccc}
\quad\includegraphics[scale=.48195]{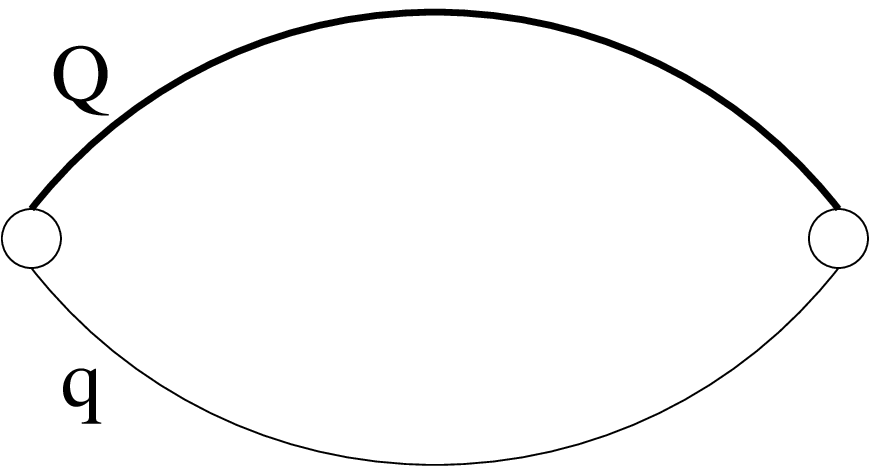}\quad&
\quad\includegraphics[scale=.48195]{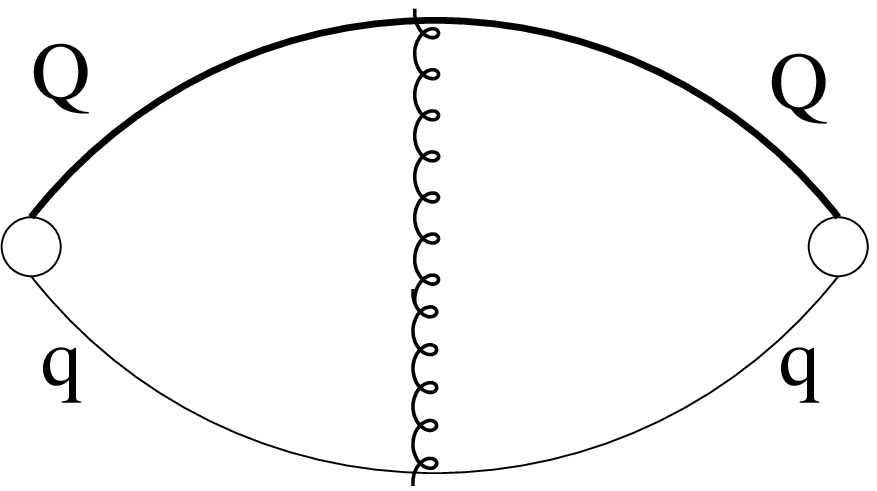}\quad&
\quad\includegraphics[scale=.48195]{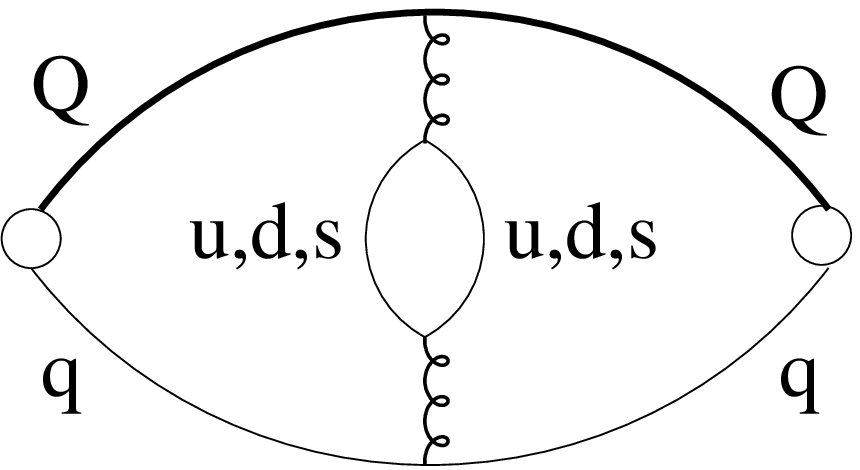}\quad\\[1ex]
(a)&(b)&(c)\end{tabular}\caption{Two-point $(Q\bar q)$ correlator:
leading (a), next-to-leading (b), next-to-next-to-leading (c)
order in~$\alpha_{\rm s}$.}\label{Fig:2}\end{center}\end{figure}

We may look at this manifestation of isospin breaking from a
related but slightly different angle \cite{LMSLD}, namely, by
applying the QCD sum-rule formalism sketched above in the
so-called local-duality limit,\footnote{This limit of QCD sum
rules has been applied to pion and nucleon elastic and meson
transition form factors \cite{LD0,BLM}.} realized by the Borel
variable approaching its lower boundary, that is, by $\tau\to0$,
to Borelized correlation functions of mass dimension two,
corresponding to exponent $N=0$. In this \emph{well-defined\/}
limit, the power corrections $\Pi^{(N)}$ vanish, the QCD sides of
the emerging QCD sum rules simplify to dispersion integrals of the
spectral densities $\rho(s,m_Q,m_q,\alpha_{\rm s}\mid m_{\rm
sea})$ (indicating also their dependence\pagebreak\ on the masses
$m_{\rm sea}$ of all sea quarks showing up in higher-order
corrections), and all nonperturbative effects are incorporated by
their upper limits of integration, the effective~thresholds
$s_{\rm eff}(m_Q,m_q,\alpha_{\rm s})$:\begin{equation}f_{H_q}^2=
\hspace{-2.3ex} \int\limits_{(m_Q+m_q)^2}^{s_{\rm
eff}(m_Q,m_q,\alpha_{\rm s})}\hspace{-2.3ex}{\rm
d}s\,\rho(s,m_Q,m_q,\alpha_{\rm s}\mid m_{\rm
sea})\equiv\digamma(s_{\rm eff}(m_q,\cdot),m_q\mid m_{\rm sea})\
.\label{f}\end{equation}For simplicity of notation, we denote the
QCD side by $\digamma(s_{\rm eff}(m_q,\cdot),m_q\mid m_{\rm sea})$
and highlight only its dependence on the light-quark masses. At
the hadron side, in the local-duality limit any dependence on the
mass of the ground state disappears, whose footprint is reduced to
its decay constant squared. In the spectral densities'
perturbative expansions (Fig.~\ref{Fig:2}), sea quarks begin to
contribute at order~$\alpha_{\rm s}^2$:
$$\rho(s,m_Q,m_q,\alpha_{\rm s}\mid m_{\rm sea})=\rho_0(s,m_Q,m_q)
+\mbox{$\frac{\alpha_{\rm s}(\mu)}{\pi}$}\,\rho_1(s,m_Q,m_q,\mu)
+\mbox{$\frac{\alpha_{\rm s}^2(\mu)}{\pi^2}$}\,
\rho_2(s,m_Q,m_q,\mu\mid m_{\rm sea})+\cdots\ .$$Truncations of
such perturbative series lead to unphysical dependences on
renormalization scales~$\mu$.

The heavy--light spectral densities required as input have been
computed up to order $\alpha_{\rm s}^2$~\cite{SD};~at order
$\alpha_{\rm s}^2$, however, only for the case of massless light
quarks: $\rho_2(s,m_Q,m_q\mid m_{\rm
sea})\approx\rho_2(s,m_Q,0\mid 0)$.

\begin{figure}[b]\begin{center}
\includegraphics[scale=.46756]{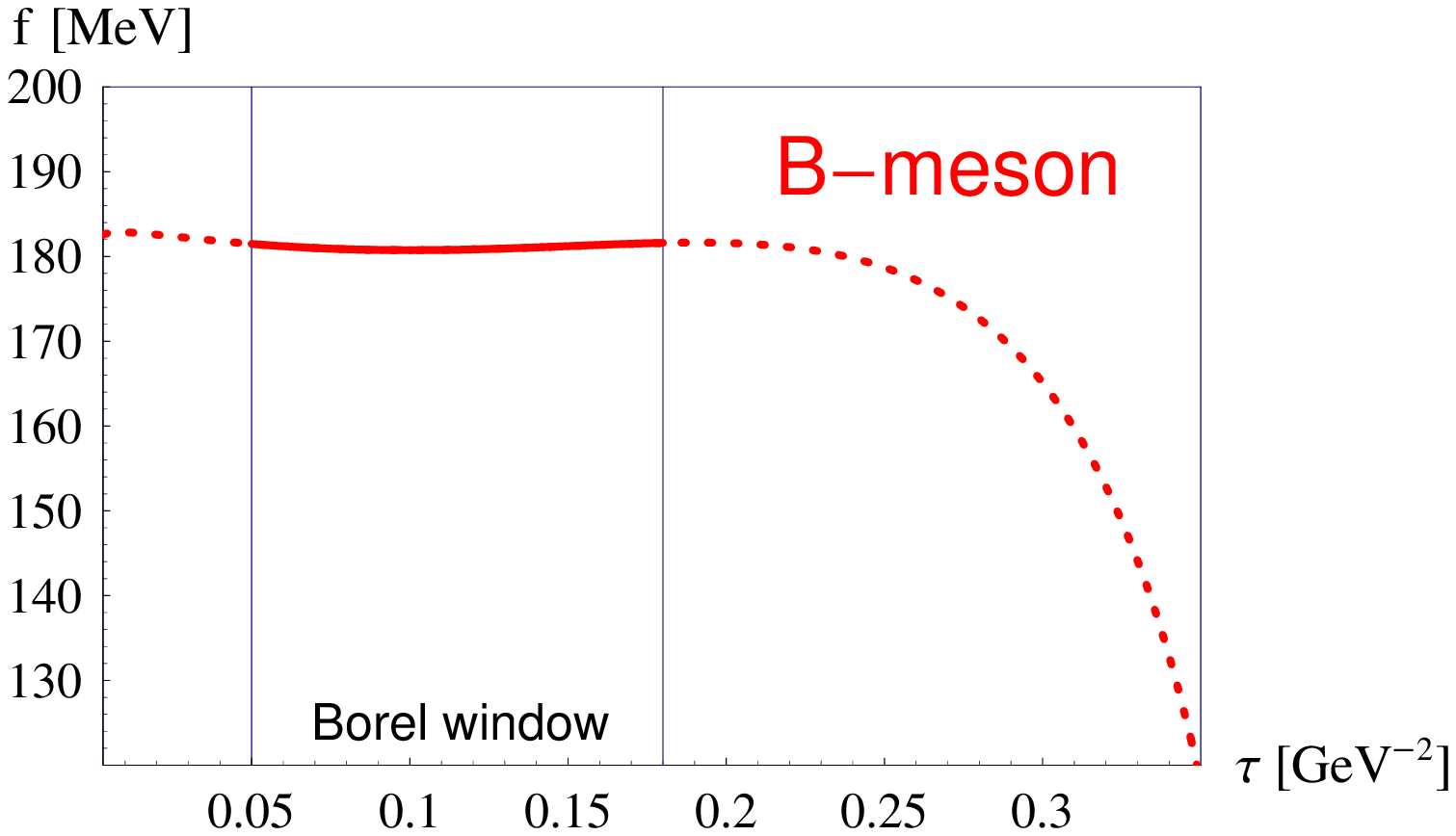}\qquad
\includegraphics[scale=.46756]{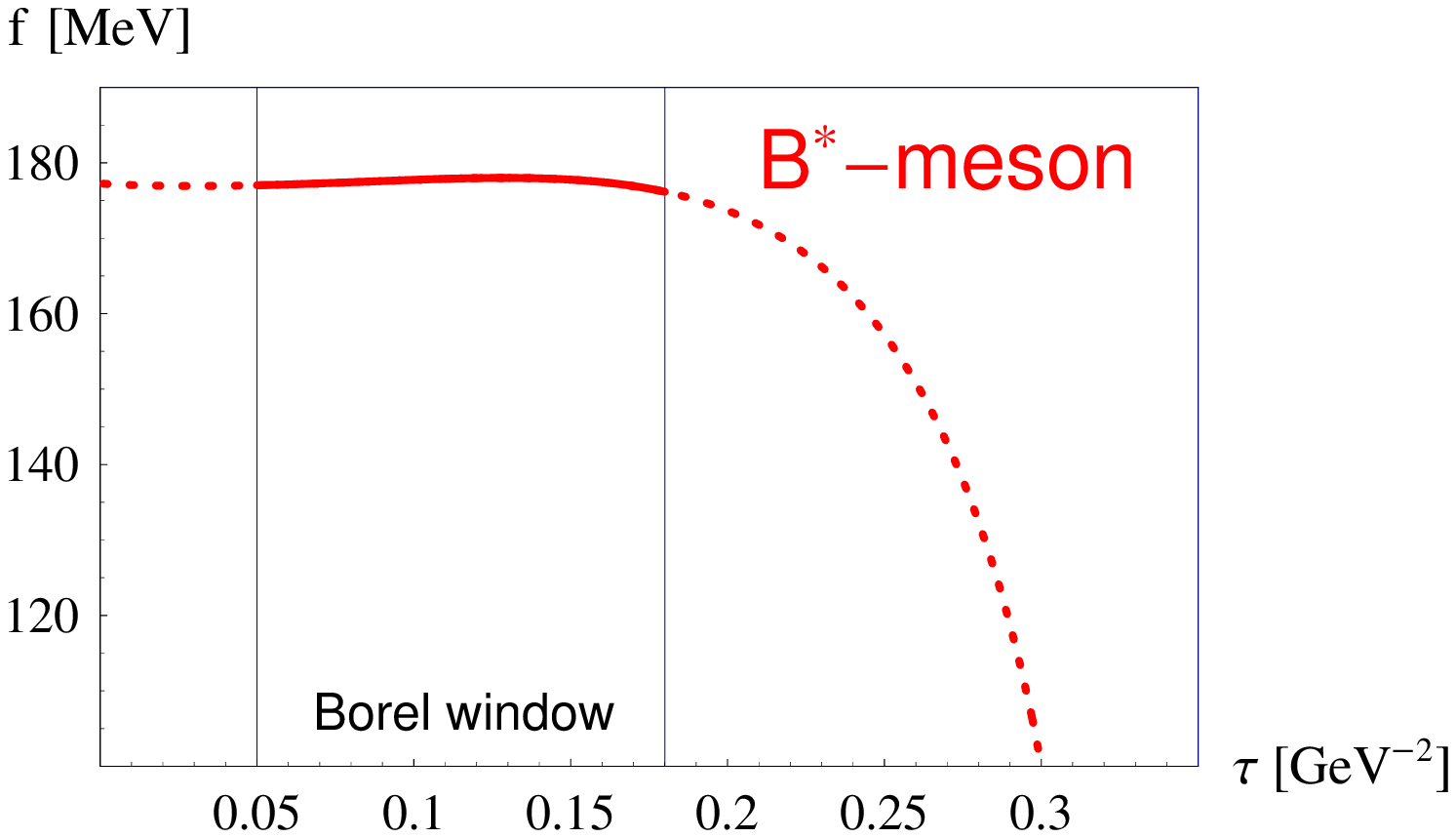}
\caption{Predictions of Borel QCD sum rules for the $B$- and
$B^*$-meson decay constants \cite{LMSLD} vs.\ Borel~variable
$\tau$, for one and the same effective threshold $s_{\rm eff}$
within (solid line) and beyond (dotted line) the Borel windows.
For both pseudoscalar and vector mesons, demanding Borel stability
over the windows fixes the values of~$s_{\rm eff}$.}\label{Fig:LD}
\end{center}\end{figure}

It is easy to convince oneself that adopting this local-duality
limit $\tau\to0$ is both mathematically well-defined and
physically well-grounded: Traditional Borel QCD sum rules identify
constants $s_{\rm eff}$ such that within chosen Borel windows the
predicted hadron observables exhibit the least sensitivity to the
value of $\tau$. However, Fig.~\ref{Fig:LD} illustrates that these
regions can safely be extended down to~$\tau=0$; the effective
thresholds found by requiring Borel stability guarantee the latter
to hold also for $\tau\to0$.

\begin{figure}[b]\begin{center}
\includegraphics[scale=.39417]{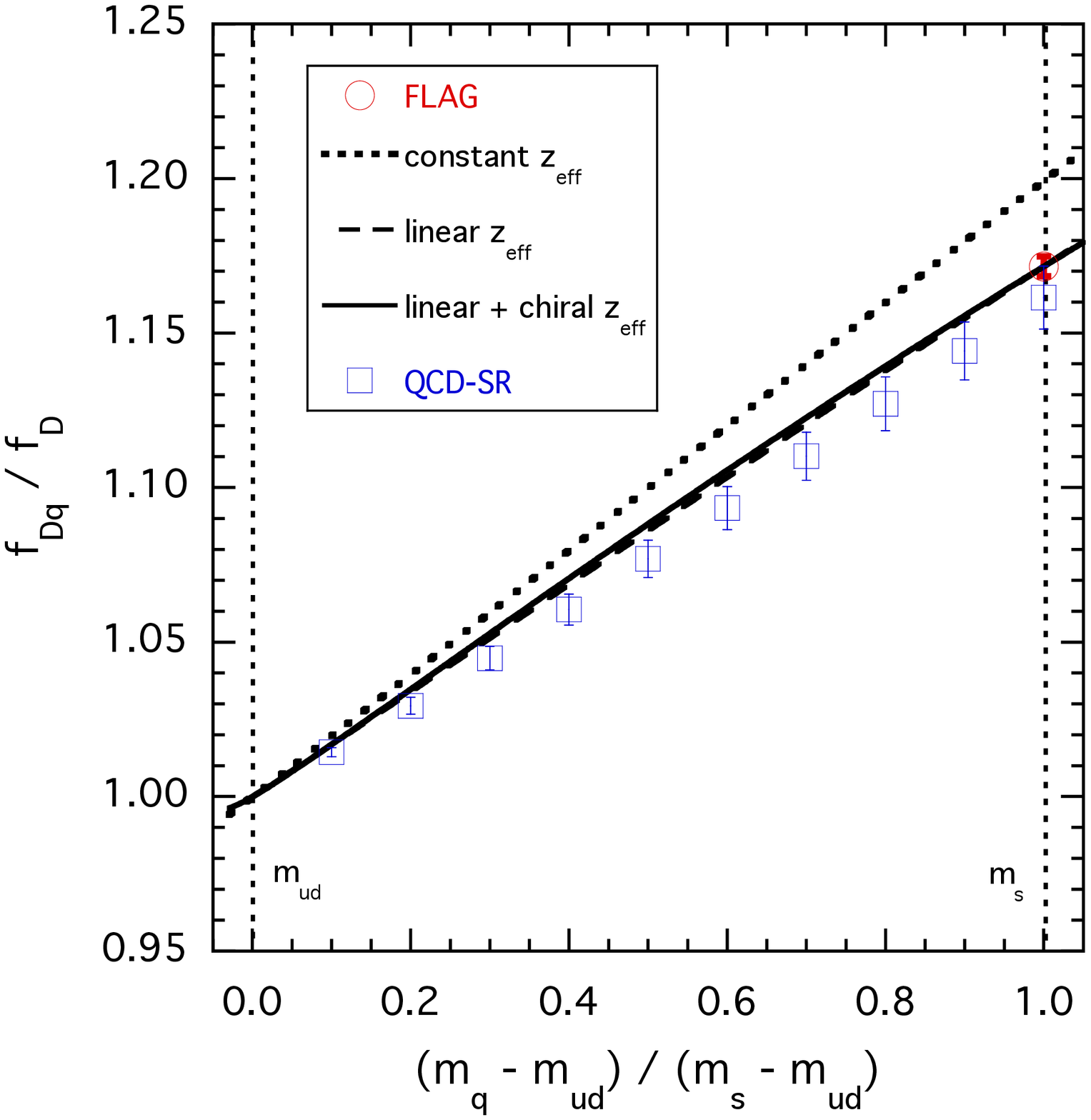}\qquad
\includegraphics[scale=.39417]{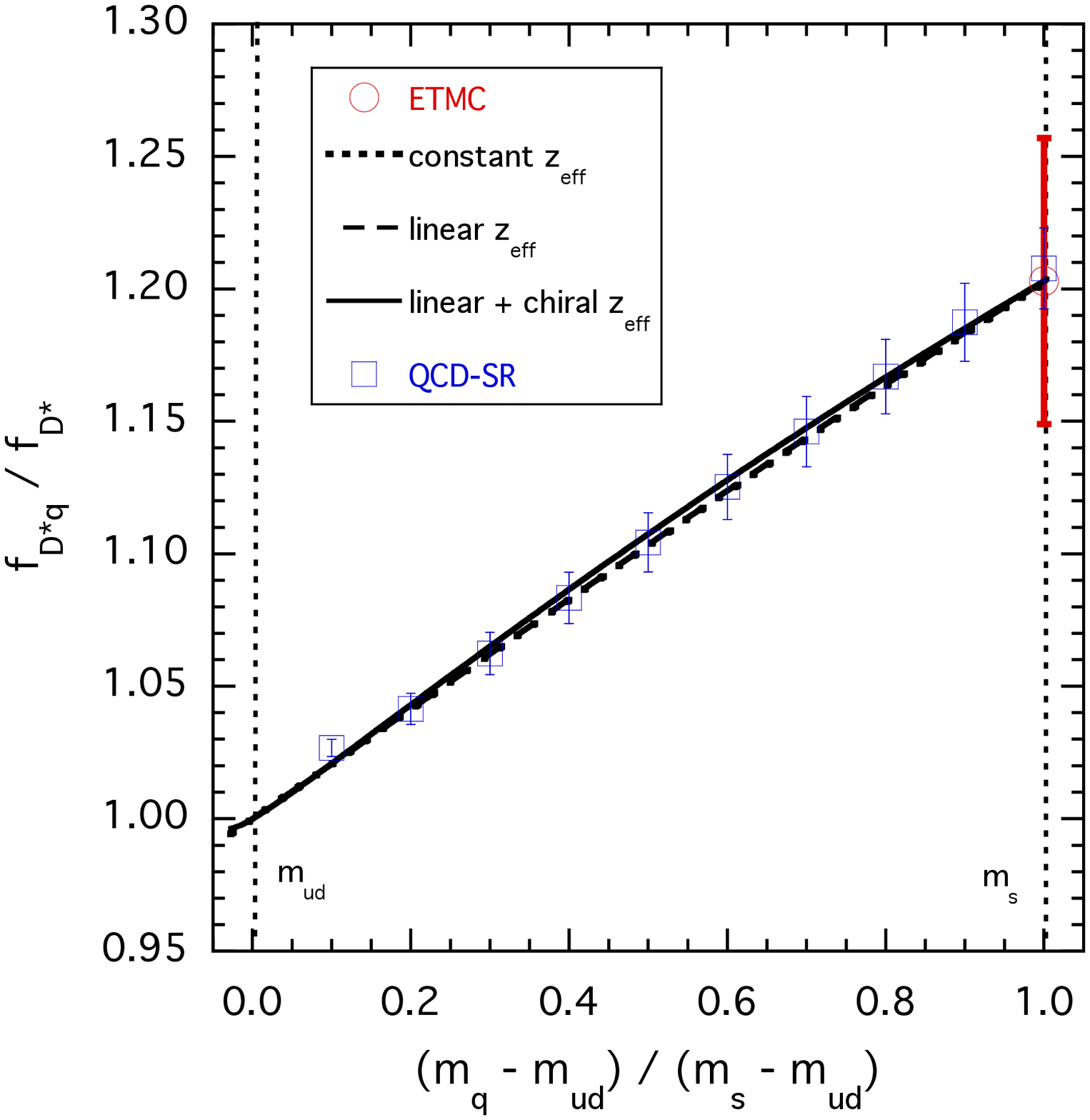}\\[3ex]
\includegraphics[scale=.39417]{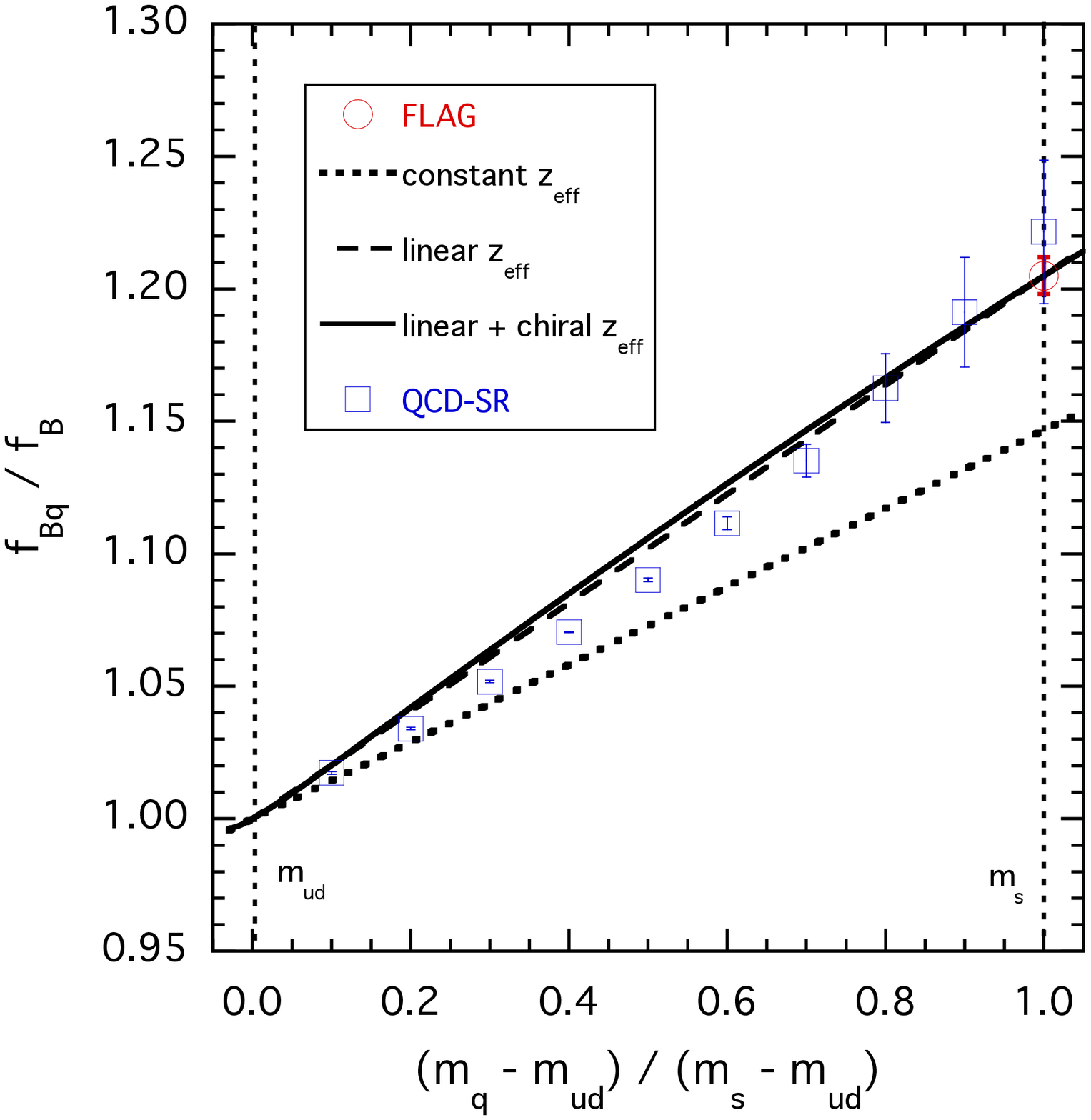}\qquad
\includegraphics[scale=.39417]{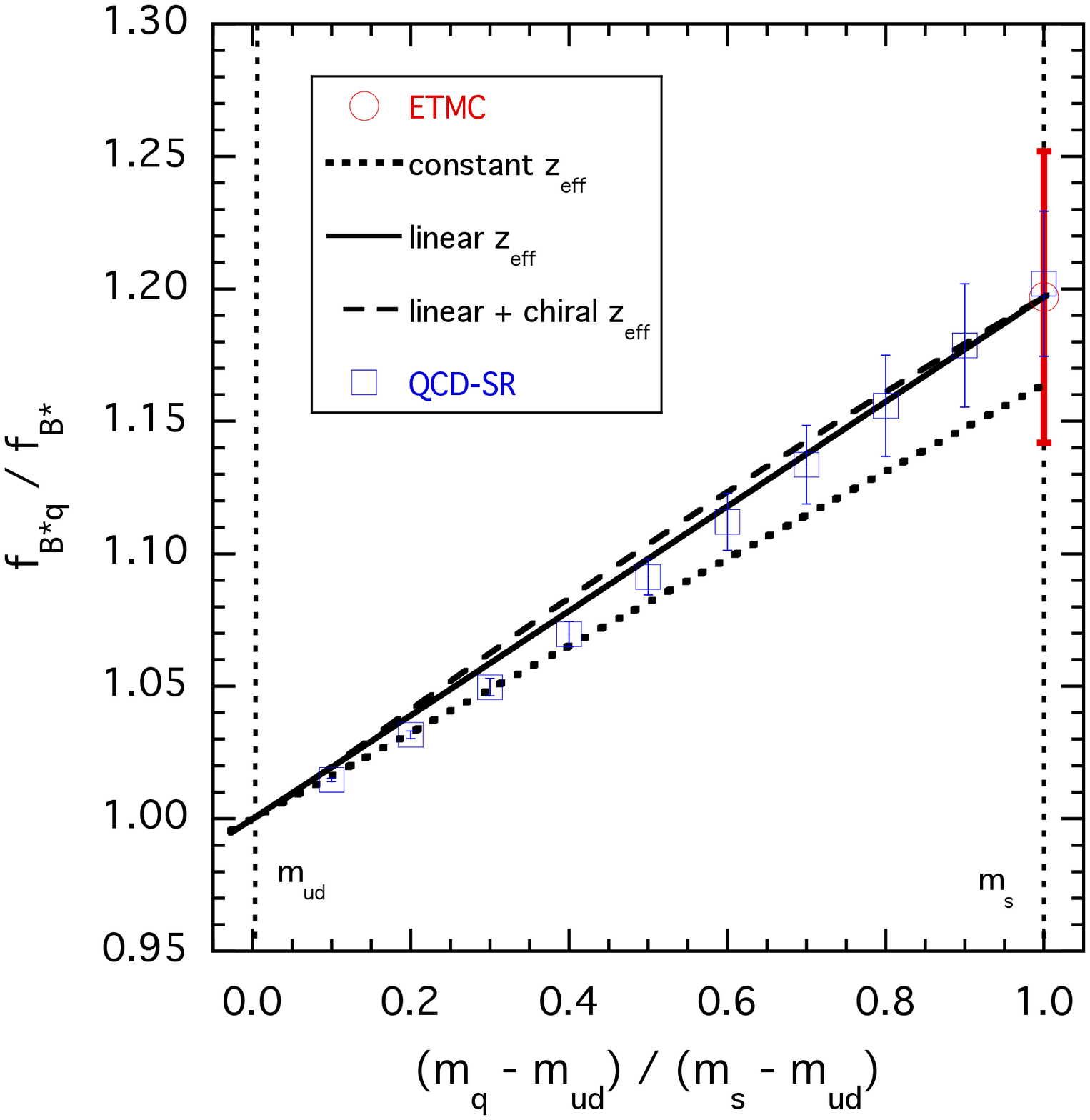}
\caption{Decay-constant ratio $f_{H_q}(m_q)/f_{H_q}(m_{ud}),$
$m_{ud}\equiv\frac{1}{2}\,(m_u+m_d),$ as a function of the
light-quark-mass variable $(m_q-m_{ud})/(m_s-m_{ud})$ for three
different ans\"atze (constant, linear, and linear plus chiral
logarithms) for redefined effective threshold $z_{\rm
eff}\equiv\sqrt{s_{\rm eff}}-m_Q-m_q$ \cite{LMSLD}, compared with
the findings (\textcolor{blue}{squares}) of Ref.~\cite{LMSIB}.}
\label{Fig:f}\end{center}\end{figure}

\section{Leptonic Decay Constant of Heavy--Light Meson: Light-Quark
Mass Dependence}We study the leptonic decay constants, $f_{P_q}$
and $f_{V_q}$, of pseudoscalar ($P_q$) and vector ($V_q$) mesons
with masses $M_{P_q}$ and $M_{V_q}$, four-momentum $p$ and $V_q$
polarization vector $\varepsilon_\mu(p)$, defined
according~to$$\langle0|\,\bar q(0)\,\gamma_\mu\,\gamma_5\,Q(0)\,
|P_q(p)\rangle={\rm i}\,f_{P_q} \,p_\mu\ ,\qquad\langle0|\,\bar
q(0)\,\gamma_\mu\,Q(0)\,|V_q(p)\rangle=f_{V_q}\,M_{V_q}\,
\varepsilon_\mu(p)\ .$$The decay-constant discrepancy in the focus
of our interest, $f_{H_d}-f_{H_u}$, betraying isospin breaking, is
proportional to the difference of the spectral integral
$\digamma(s_{\rm eff}(m_q,\cdot),m_q\mid m_{\rm sea})$ in
Eq.~(\ref{f}) if evaluated at the light-quark masses $m_q=m_d$ and
$m_q=m_u$; in that difference, all dependence on the strange sea
quarks cancels, reducing greatly the uncertainties induced by
neglect of $m_{\rm sea}$ in $\rho_2(s,m_Q,m_q\mid m_{\rm sea})$:
\begin{align*}\delta\digamma &\equiv\digamma(s_{\rm
eff}(m_d),m_d\mid m_{\rm sea})-\digamma(s_{\rm eff}(m_u),m_u\mid
m_{\rm sea})\\&=\digamma(s_{\rm eff}(m_d),m_d\mid m_{\rm
sea}=0)-\digamma(s_{\rm eff}(m_u),m_u\mid m_{\rm sea}=0)
+O\!\left(\frac{\alpha_{\rm s}^2}{\pi^2}\,(m_d-m_u)\right).
\end{align*}The dependence of $\digamma(s_{\rm
eff}(m_q,\cdot),m_q\mid m_{\rm sea})$ on $m_q$ originates
basically in two quantities therein: The one of the spectral
densities we easily extract by adapting results available in the
literature \cite{SD}. The one of the effective thresholds $s_{\rm
eff}(m_q)=s_0+s_1\,m_q+\cdots$ we derive by allowing the
light-quark mass $m_q$ to vary continuously between chiral limit,
$m_q=0$, and strange-quark mass, $m_q=m_s$, defining an
$H_q$-meson decay constant function $f_H(m_q)$ by the emerging
outcomes of the local-duality QCD sum rule (\ref{f}) for
$m_q\in[0,m_s]$, and matching (cf.~Fig.~\ref{Fig:f}) the behaviour
of $f_H(m_q)$ to lattice-QCD findings \cite{LQCD} for
$f_H((m_u+m_d)/2)$ and $f_H(m_s)$. From the hence fully
determined~$m_q$~dependence of the QCD sum-rule outcome and
numerical values of all QCD parameters in the modified
minimal-subtraction renormalization scheme, we predict, for the
$D$, $D^*$, $B$ and $B^*$ mesons, the decay-constant~differences
\begin{align*}f_{D^\pm}-f_{D^0}&=(0.96\pm0.09)\;\mbox{MeV}\ ,&
f_{D^{*\pm}}-f_{D^{*0}}&=(1.18\pm0.35)\;\mbox{MeV}\ ,\\[1ex]
f_{B^0}-f_{B^\pm}&=(1.01\pm0.10)\;\mbox{MeV}\ ,&
f_{B^{*0}}-f_{B^{*\pm}}&=(0.89\pm0.30)\;\mbox{MeV}\ .\end{align*}
The proximity of the $f_H(m_q)$ curves resulting from constant
(\emph{i.e.}, $m_q$-independent) and non-constant (\emph{i.e.},
$m_q$-dependent) parametrizations of $z_{\rm eff}\equiv
\sqrt{s_{\rm eff}}-m_Q-m_q$ in Fig.~\ref{Fig:f} shows that roughly
$70\mbox{--}80\%$ of the strong isospin breaking in decay
constants is due to our spectral densities' --- analytically and
rigorously derivable --- $m_q$ behaviour enabling us to retain
control over the accuracy of our~findings.

\acknowledgments{D.~M.~acknowledges support by the Austrian
Science Fund (FWF) under Project P 29028-N27.}

\end{document}